\PassOptionsToPackage{unicode}{hyperref}
\PassOptionsToPackage{hyphens}{url}
\documentclass[
]{article}
\usepackage{amsmath,amssymb}
\usepackage{lmodern}
\usepackage{ifxetex,ifluatex}
\ifnum 0\ifxetex 1\fi\ifluatex 1\fi=0 
  \usepackage[T1]{fontenc}
  \usepackage[utf8]{inputenc}
  \usepackage{textcomp} 
\else 
  \usepackage{unicode-math}
  \defaultfontfeatures{Scale=MatchLowercase}
  \defaultfontfeatures[\rmfamily]{Ligatures=TeX,Scale=1}
\fi
\IfFileExists{upquote.sty}{\usepackage{upquote}}{}
\IfFileExists{microtype.sty}{
  \usepackage[]{microtype}
  \UseMicrotypeSet[protrusion]{basicmath} 
}{}
\makeatletter
\@ifundefined{KOMAClassName}{
  \IfFileExists{parskip.sty}{%
  }{
    \setlength{\parindent}{0pt}
    \setlength{\parskip}{6pt plus 2pt minus 1pt}}
}{
  \KOMAoptions{parskip=half}}
\makeatother
\usepackage{xcolor}
\IfFileExists{xurl.sty}{\usepackage{xurl}}{} 
\IfFileExists{bookmark.sty}{\usepackage{bookmark}}{\usepackage{hyperref}}
\hypersetup{
  pdftitle={On rank statistics of PageRank and MarkovRank},
  pdfauthor={Yoichi Nishiyama},
  hidelinks,
  pdfcreator={LaTeX via pandoc}}
\usepackage[margin=1in]{geometry}
\usepackage{color}
\usepackage{fancyvrb}

\DefineVerbatimEnvironment{Highlighting}{Verbatim}{commandchars=\\\{\}}
\usepackage{framed}
\definecolor{shadecolor}{RGB}{248,248,248}
\newenvironment{Shaded}{\begin{snugshade}}{\end{snugshade}}

\newcommand{\AttributeTok}[1]{\textcolor[rgb]{0.77,0.63,0.00}{#1}}

\newcommand{\CommentTok}[1]{\textcolor[rgb]{0.56,0.35,0.01}{\textit{#1}}}

\newcommand{\ConstantTok}[1]{\textcolor[rgb]{0.00,0.00,0.00}{#1}}
\newcommand{\ControlFlowTok}[1]{\textcolor[rgb]{0.13,0.29,0.53}{\textbf{#1}}}

\newcommand{\DecValTok}[1]{\textcolor[rgb]{0.00,0.00,0.81}{#1}}

\newcommand{\FloatTok}[1]{\textcolor[rgb]{0.00,0.00,0.81}{#1}}
\newcommand{\FunctionTok}[1]{\textcolor[rgb]{0.00,0.00,0.00}{#1}}

\newcommand{\NormalTok}[1]{#1}

\newcommand{\OtherTok}[1]{\textcolor[rgb]{0.56,0.35,0.01}{#1}}

\newcommand{\SpecialCharTok}[1]{\textcolor[rgb]{0.00,0.00,0.00}{#1}}

\newcommand{\StringTok}[1]{\textcolor[rgb]{0.31,0.60,0.02}{#1}}

\usepackage{graphicx}
\makeatletter
\def\maxwidth{\ifdim\Gin@nat@width>\linewidth\linewidth\else\Gin@nat@width\fi}
\def\maxheight{\ifdim\Gin@nat@height>\textheight\textheight\else\Gin@nat@height\fi}
\makeatother
\setkeys{Gin}{width=\maxwidth,height=\maxheight,keepaspectratio}
\makeatletter
\def\fps@figure{htbp}
\makeatother
\providecommand{\tightlist}{%
  \setlength{\itemsep}{0pt}\setlength{\parskip}{0pt}}
\ifluatex
  \usepackage{selnolig}  
\fi

\title{On rank statistics of \emph{PageRank} and \emph{MarkovRank}}
\author{Yoichi Nishiyama \\ {\sc\normalsize Waseda University}}
\date{June 1, 2022}

\begin{document}
\maketitle

\begin{abstract}
The well-known statistic \emph{PageRank} was created in 1998 by
co-founders of Google, Sergey Brin and Larry Page, to optimize the
ranking of websites for their search engine outcomes. It is computed
using an iterative algorithm, based on the idea that nodes with a larger
number of incoming edges are more important. Google's PageRank involves 
some information from ``aliens''; the 15\% of information is regarded 
as the connections from the outside of the network system under consideration. 

In this paper, seeking a stable statistic which is ``close'' to an
``intrinsic'' version of PageRank, we will introduce a new statistic
called \emph{MarkovRank}. A special attention will be paid to the
comparison of \emph{rank statistics} among standard-PageRank,
``intrinsic-PageRank'' and MarkovRank. It is concluded that the
rank statistic of MarkovRank, which is always well-defined, is identical
to that of ``intrinsic-PageRank'', as far as the latter is well-defined.
\end{abstract}

\hypertarget{introduction}{%
\section{Introduction}\label{introduction}}

The well-known statistic \emph{PageRank} was created in 1998 by
co-founders of Google, Sergey Brin and Larry Page, to optimize the
ranking of websites for their search engine outcomes. It is computed
using an iterative algorithm, based on the idea that nodes with a larger
number of incoming edges are more important.

Google's PageRank involves some information from ``aliens''; the 15\% of
information is regarded as the connections from the outside of the
network system under consideration. Without involving the information
from ``aliens'', Google's PageRank could not be well-defined.

In this paper, seeking a stable statistic which is ``close'' to an
``intrinsic'' version of PageRank, we will introduce a new statistic
called \emph{MarkovRank}. A special attention will be paid to the
comparison of \emph{rank statistics} among standard-PageRank,
``intrinsic-PageRank'' and MarkovRank, and it is concluded that the
rank statistic of MarkovRank, which is always well-defined, is identical
to that of ``intrinsic-PageRank'', as far as the latter is well-defined.
Notice that, since the sum of either ``Rank'' values across all nodes
equals \(1\), the absolute values of the ``Rank''s in each of concrete
network data are not meaningful, and therefore the rank statistics of
them are important. Since it is confirmed by some examples that
MarkovRank returns as stable values as standard-PageRank, it can be said
that MarkovRank has a potential to play a similar role to the well
established standard-PageRank, not only from a practical point of view
but also with some new theoretical validity. Although there are huge
literature dealing with PageRank and its variations, including Bianchini
\emph{et al.} (2005), Bar-Yossef and Mashiach (2008) and Langville and
Meyer (2011), the approach presented in the current paper seems novel.

To close this section, let us load two packages in R.

\begin{Shaded}
\begin{Highlighting}[]
\FunctionTok{library}\NormalTok{(}\StringTok{"MASS"}\NormalTok{)   }\CommentTok{\# for the calculation of generalized inverse}
\FunctionTok{library}\NormalTok{(}\StringTok{"igraph"}\NormalTok{) }\CommentTok{\# for the analysis of graphical data}
\end{Highlighting}
\end{Shaded}

\hypertarget{preliminary-disucssions}{%
\section{Preliminary disucssions}\label{preliminary-disucssions}}

\hypertarget{an-ideal-pagerank-for-the-easiest-case}{%
\subsection{An ``ideal-PageRank'' (for the easiest
case only)}\label{an-ideal-pagerank-for-the-easiest-case}}

First let us observe the following network data for an illustration.

\begin{Shaded}
\begin{Highlighting}[]
\NormalTok{(A }\OtherTok{\textless{}{-}} \FunctionTok{matrix}\NormalTok{(}\FunctionTok{c}\NormalTok{(}\DecValTok{0}\NormalTok{,}\DecValTok{1}\NormalTok{,}\DecValTok{0}\NormalTok{,}\DecValTok{0}\NormalTok{,}\DecValTok{1}\NormalTok{,}\DecValTok{0}\NormalTok{,}\DecValTok{1}\NormalTok{,}\DecValTok{1}\NormalTok{,}\DecValTok{0}\NormalTok{,}\DecValTok{1}\NormalTok{,}\DecValTok{0}\NormalTok{,}\DecValTok{0}\NormalTok{,}\DecValTok{1}\NormalTok{,}\DecValTok{0}\NormalTok{,}\DecValTok{0}\NormalTok{,}\DecValTok{0}\NormalTok{), }\AttributeTok{nrow=}\DecValTok{4}\NormalTok{))}
\end{Highlighting}
\end{Shaded}

\begin{verbatim}
##      [,1] [,2] [,3] [,4]
## [1,]    0    1    0    1
## [2,]    1    0    1    0
## [3,]    0    1    0    0
## [4,]    0    1    0    0
\end{verbatim}

\includegraphics{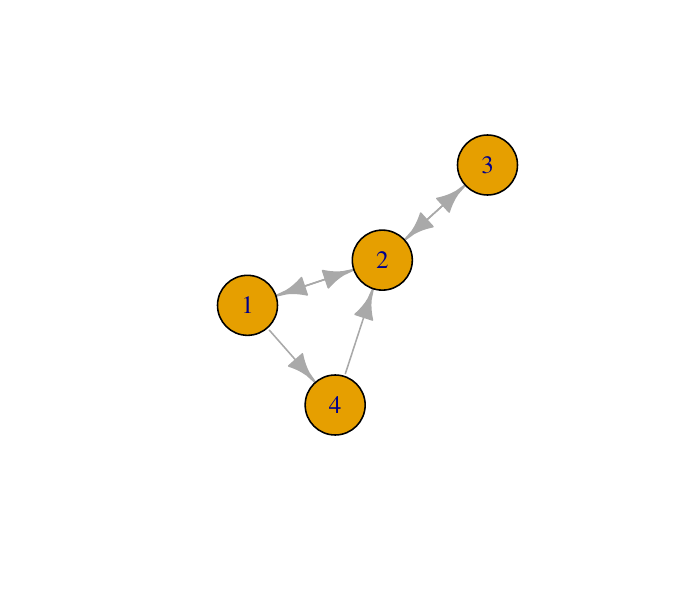}

An intuitive interpretation of the above network is the following:

\begin{itemize}
\tightlist
\item
  The person \(1\) follows the tweets of the persons \(2\) and \(4\);
\item
  The person \(2\) does those of the persons \(1\) and \(3\);
\item
  The person \(3\) does those of the person \(2\);
\item
  The person \(4\) does those of the person \(2\).
\end{itemize}

Such information of the network is contained in the \emph{adjacency
matrix} \(A\) given by \[A=\left(\begin{array}{cccc} 0 & 1 & 0 & 1\\
1 & 0 & 1 & 0\\
0 & 1 & 0 & 0\\
0 & 1 & 0 & 0\end{array}\right),\] where \[A_{ij}=\left\{
\begin{array}{cc}1, & \quad \mbox{if the person $i$ follows the tweets of the person $j$}, \\ 0, & \quad \mbox{otherwise}. \end{array}\right.\]

Looking at each column of the matrix \(A\), we find that:

\begin{itemize}
\tightlist
\item
  the person \(2\) is followed by three persons;
\item
  the others are followed only by one person;
\item
  in particular, the persons \(1\) and \(3\) are followed by the person
  \(2\), and the person \(4\) is not.
\end{itemize}

Thus, we can easily imagine that the Twitter account of the person \(2\)
is the most popular among all, and that those of the persons \(1\) and
\(3\) are more popular than that of \(4\).

Now, let \(B\) be the diagonal matrix whose \((i,i)\) entry is the
out-degree of the node \(i\); that is,
\[B=\left(\begin{array}{cccc} 2 & 0 & 0 & 0\\
0 & 2 & 0 & 0\\
0 & 0 & 1 & 0\\
0 & 0 & 0 & 1\end{array}\right).\] Then, the matrix
\begin{eqnarray*}M=A^TB^{-1}&=&\left(\begin{array}{cccc} 0 & 1 & 0 & 0\\
1 & 0 & 1 & 1\\
0 & 1 & 0 & 0\\
1 & 0 & 0 & 0\end{array}\right)
\left(\begin{array}{cccc} 1/2 & 0 & 0 & 0\\
0 & 1/2 & 0 & 0\\
0 & 0 & 1 & 0\\
0 & 0 & 0 & 1\end{array}\right)
\\
&=&
\left(\begin{array}{cccc} 0 & 1/2 & 0 & 0\\
1/2 & 0 & 1 & 1\\
0 & 1/2 & 0 & 0\\
1/2 & 0 & 0 & 0\end{array}\right)\end{eqnarray*} is a transition one;
actually, all entries are non-negative and the sum of each column is
\(1\).

Note that, in this \emph{easy} case:

\begin{itemize}
\tightlist
\item
  \(M\) is (already) the transition matrix of a Markov chain;
\item
  The Markov chain is regular (thus \(M\) has a unique eigenvalue
  \(1\)).
\end{itemize}

\par \vskip 5pt \noindent
\textbf{Definition 1. {[}ideal-PageRank{]}} For given \((n \times n)\)
adjacency matrix \(A\), construct the matrix \(M\) with the procedure
described above. When this \(M\) becomes the transition matrix of a
regular Markov chain, the ``ideal-PageRank'' is defined as the
\((n \times 1)\) vector \({\bf p}_\infty\) given by
\[{\bf p}_\infty=\lim_{k \to \infty}M^k{\bf p}_0,\] with some (actually,
any) initial value \({\bf p}_0=(p_{0}^{(1)},...,p_{0}^{(n)})^T\) such
that \(p_{0}^{(i)}\geq 0\) for all \(i\) and that
\(\sum_{i=1}^n p_{0}^{(i)}=1\).
\par \vskip 5pt \noindent

Although the mathematical calculation of the ideal-PageRank via the
diagonalization of \(M\) is possible, a numerical computation based on
the recurrence equation is faster especially when \(n\) is large.

\begin{Shaded}
\begin{Highlighting}[]
\NormalTok{MC\_limit }\OtherTok{\textless{}{-}} \ControlFlowTok{function}\NormalTok{(M)\{}
\NormalTok{  n }\OtherTok{\textless{}{-}} \FunctionTok{nrow}\NormalTok{(M)}
\NormalTok{  p }\OtherTok{\textless{}{-}} \FunctionTok{matrix}\NormalTok{(}\DecValTok{1}\SpecialCharTok{/}\NormalTok{n, }\AttributeTok{nrow=}\NormalTok{n, }\AttributeTok{ncol=}\DecValTok{1}\NormalTok{)}
\NormalTok{  diff }\OtherTok{\textless{}{-}} \DecValTok{1}
  \ControlFlowTok{while}\NormalTok{(diff }\SpecialCharTok{\textgreater{}} \FloatTok{0.0000001}\NormalTok{)\{}
\NormalTok{    p.pre }\OtherTok{\textless{}{-}}\NormalTok{ p}
\NormalTok{    p }\OtherTok{\textless{}{-}}\NormalTok{ M }\SpecialCharTok{\%*\%}\NormalTok{ p}
\NormalTok{    diff }\OtherTok{\textless{}{-}} \FunctionTok{max}\NormalTok{(}\FunctionTok{abs}\NormalTok{(p}\SpecialCharTok{{-}}\NormalTok{p.pre))}
\NormalTok{  \}}
  \FunctionTok{return}\NormalTok{(}\FunctionTok{t}\NormalTok{(p))}
\NormalTok{\}}
\end{Highlighting}
\end{Shaded}

We are now ready to calculate the ideal-PageRank for the network data
given above.

\begin{Shaded}
\begin{Highlighting}[]
\NormalTok{n }\OtherTok{\textless{}{-}} \FunctionTok{nrow}\NormalTok{(A)}
\NormalTok{M }\OtherTok{\textless{}{-}} \FunctionTok{t}\NormalTok{(A) }\SpecialCharTok{\%*\%} \FunctionTok{diag}\NormalTok{(}\DecValTok{1}\SpecialCharTok{/}\FunctionTok{rowSums}\NormalTok{(A), }\AttributeTok{nrow=}\NormalTok{n)}
\FunctionTok{MC\_limit}\NormalTok{(M)}
\end{Highlighting}
\end{Shaded}

\begin{verbatim}
##           [,1]      [,2]      [,3]      [,4]
## [1,] 0.2222222 0.4444444 0.2222222 0.1111111
\end{verbatim}

\hypertarget{an-intrinsic-pagerank-for-the-regular-case-only}{%
\subsection{An ``intrinsic-PageRank'' (for the regular case
only)}\label{an-intrinsic-pagerank-for-the-regular-case-only}}

In general, however, since some of the out-degrees of the adjacency
matrix \(A\) may be zero (i.e., some person may not follow any other
person's tweets), the matrix \(B\) may not be invertible; thus we may
not be able to introduce the transition matrix ``\(M=A^TB^{-1}\)''.

Unfortunately, the idea just replacing \(B^{-1}\) with the generalized
inverse \(B^{-}\) is not sufficient, because \(A^T B^{-}\) is still not
a transition matrix.

\par \vskip 5pt \noindent
\textbf{Definition 2. {[}The transition matrix \(\widetilde{M}\){]}} For
any given \((n \times n)\) adjacency matrix \(A\), the corresponding
transition matrix \(\widetilde{M}\), which coincides with \(M\) if \(B\)
is invertible, is defined by
\[\widetilde{M}=A^TB^{-}+\frac{1}{n}{\bf 1}(I-BB^{-}),\] where
\({\bf 1}\) denotes the \((n \times n)\) matrix whose all entries are
\(1\).
\par \vskip 5pt \noindent

To get better understanding of the above definition of
\(\widetilde{M}\), let us observe a concrete example. 
If unfortunately the matrix \(B\) is not invertible, like
\[A=\left(\begin{array}{cccc} 0 & 1 & 0 & 1\\
1 & 0 & 1 & 0\\
0 & 1 & 0 & 0\\
0 & 0 & 0 & 0\end{array}\right)\quad \mbox{and} \quad
B=\left(\begin{array}{cccc} 2 & 0 & 0 & 0\\
0 & 2 & 0 & 0\\
0 & 0 & 1 & 0\\
0 & 0 & 0 & 0\end{array}\right),\] since the matrix \(A^TB^{-}\), such
as \[\left(\begin{array}{cccc} 0 & 1 & 0 & 0\\
1 & 0 & 1 & 0\\
0 & 1 & 0 & 0\\
1 & 0 & 0 & 0\end{array}\right)
\left(\begin{array}{cccc} 1/2 & 0 & 0 & 0\\
0 & 1/2 & 0 & 0\\
0 & 0 & 1 & 0\\
0 & 0 & 0 & 0\end{array}\right)
=
\left(\begin{array}{cccc} 0 & 1/2 & 0 & 0\\
1/2 & 0 & 1 & 0\\
0 & 1/2 & 0 & 0\\
1/2 & 0 & 0 & 0\end{array}\right),\] is not a transition one, then the
matrix \(\frac{1}{n}{\bf 1}(I-BB^{-})\) should be added to \(A^TB^{-}\)
to obtain the transition matrix \(\widetilde{M}\); that is,
\begin{eqnarray*}\widetilde{M}&=&A^T B^{-}+\frac{1}{n}{\bf 1}(I-BB^{-})
\\
&=&
\left(\begin{array}{cccc} 0 & 1/2 & 0 & 0\\
1/2 & 0 & 1 & 0\\
0 & 1/2 & 0 & 0\\
1/2 & 0 & 0 & 0\end{array}\right)+
\left(\begin{array}{cccc} 0 & 0 & 0 & 1/4\\
0 & 0 & 0 & 1/4\\
0 & 0 & 0 & 1/4\\
0 & 0 & 0 & 1/4\end{array}\right)
\\
&=&\left(\begin{array}{cccc} 0 & 1/2 & 0 & 1/4\\
1/2 & 0 & 1 & 1/4\\
0 & 1/2 & 0 & 1/4\\
1/2 & 0 & 0 & 1/4\end{array}\right).\end{eqnarray*}

An R-function to produce the matrix \(\widetilde{M}\) is given as
follows:

\begin{Shaded}
\begin{Highlighting}[]
\NormalTok{M\_tilde }\OtherTok{\textless{}{-}} \ControlFlowTok{function}\NormalTok{(adj)\{}
\NormalTok{  n }\OtherTok{\textless{}{-}} \FunctionTok{nrow}\NormalTok{(adj)}
\NormalTok{  B }\OtherTok{\textless{}{-}} \FunctionTok{diag}\NormalTok{(}\FunctionTok{rowSums}\NormalTok{(adj), n)}
\NormalTok{  one.mat }\OtherTok{\textless{}{-}} \FunctionTok{matrix}\NormalTok{(}\DecValTok{1}\NormalTok{, }\AttributeTok{nrow=}\NormalTok{n, }\AttributeTok{ncol=}\NormalTok{n)}
\NormalTok{  C }\OtherTok{\textless{}{-}} \FunctionTok{diag}\NormalTok{((}\FunctionTok{rowSums}\NormalTok{(adj)}\SpecialCharTok{==}\DecValTok{0}\NormalTok{), n)}
\NormalTok{  Mtilde }\OtherTok{\textless{}{-}} \FunctionTok{t}\NormalTok{(adj) }\SpecialCharTok{\%*\%} \FunctionTok{ginv}\NormalTok{(B) }\SpecialCharTok{+}\NormalTok{ (}\DecValTok{1}\SpecialCharTok{/}\NormalTok{n) }\SpecialCharTok{*}\NormalTok{ one.mat }\SpecialCharTok{\%*\%}\NormalTok{ C}
  \FunctionTok{return}\NormalTok{(Mtilde)}
\NormalTok{\}}
\end{Highlighting}
\end{Shaded}

We then apply the recurrent equation
\[{\bf p}_k=\widetilde{M}{\bf p}_{k-1}=\widetilde{M}^k{\bf p}_{0}, \quad k=1,2,...,\]
with the initial value \({\bf p}_0=(1/n,...,1/n)^T\), to define and
compute an \emph{intrinsic version} of \emph{PageRank}, which is still
different from the standard-PageRank originally created by Brin
and Page (1998), given by
\[\mbox{intrinsic-PR}=\lim_{k \to \infty}{\bf p}_k=\lim_{k \to \infty}\widetilde{M}^k {\bf p}_0.\]

Actually, the intrinsic-PageRank can be calculated in the following way:

\begin{Shaded}
\begin{Highlighting}[]
\NormalTok{intrinsic\_PR }\OtherTok{\textless{}{-}} \ControlFlowTok{function}\NormalTok{(adj)\{}
  \FunctionTok{MC\_limit}\NormalTok{(}\FunctionTok{M\_tilde}\NormalTok{(adj))}
\NormalTok{\}}
\end{Highlighting}
\end{Shaded}

Now that the function \texttt{intrinsic\_PR()} to compute the
intrinsic-PageRank is ready, let us double-check the ideal-PageRank for
the particular example given in the previous subsection.

\begin{Shaded}
\begin{Highlighting}[]
\FunctionTok{intrinsic\_PR}\NormalTok{(A)}
\end{Highlighting}
\end{Shaded}

\begin{verbatim}
##           [,1]      [,2]      [,3]      [,4]
## [1,] 0.2222222 0.4444444 0.2222222 0.1111111
\end{verbatim}

On the other hand, using to the function \texttt{page.rank()} included
in the package \textbf{igraph}, the standard-PageRank is computed as
follows:

\begin{Shaded}
\begin{Highlighting}[]
\NormalTok{A.ga }\OtherTok{\textless{}{-}} \FunctionTok{graph.adjacency}\NormalTok{(A, }\AttributeTok{mode=}\StringTok{"directed"}\NormalTok{)}
\FunctionTok{page.rank}\NormalTok{(A.ga)}\SpecialCharTok{$}\NormalTok{vector}
\end{Highlighting}
\end{Shaded}

\begin{verbatim}
## [1] 0.2199138 0.4292090 0.2199138 0.1309634
\end{verbatim}

We have found that the standard-PageRank and the intrinsic-PageRank
return completely different values.

\hypertarget{the-alpha-pagerank-and-a-new-statistic-called-markovrank}{%
\section{\texorpdfstring{The \(\alpha\)-PageRank and a new statistic
called
``MarkovRank''}{The \textbackslash alpha-PageRank and a new statistic called ``MarkovRank''}}\label{the-alpha-pagerank-and-a-new-statistic-called-markovrank}}

\hypertarget{the-alpha-pagerank-for-the-general-non-regular-case}{%
\subsection{\texorpdfstring{The \(\alpha\)-PageRank (for the general,
non-regular
case)}{The \textbackslash alpha-PageRank (for the general, non-regular case)}}\label{the-alpha-pagerank-for-the-general-non-regular-case}}

Below we give the definition of \(\alpha\)-PageRank. We announce in
advance that:

\begin{itemize}
\item
  \(1\)-PageRank coincides with the intrinsic-PageRank, as far as the
  multiplicity of the eigenvalue \(1\) of \(\widetilde{M}\) is one;
\item
  \(0.85\)-PageRank coincides with the (so-called) PageRank by Brin and
  Page (1998). We have already called it ``standard-PageRank'', which is
  always well-defined. 
\end{itemize}
We remark also that the formula used to define the \(\alpha\)-PageRank below is commonly called 
\emph{damping} (see, e.g., Bar-Yossef and Mashiach (2008)).

\par \vskip 5pt \noindent
\textbf{Definition 3. {[}The \(\alpha\)-PageRank{]}} Let \(A\) be any
\((n \times n)\) adjacency matrix.

\begin{enumerate}
\def\labelenumi{(\roman{enumi})}
\item
  For \(0<\alpha < 1\), the \(\alpha\)-PageRank is (always) well-defined
  as
  \[\alpha\mbox{-PageRank}=\lim_{k\to\infty}\widehat{M}_\alpha^k{\bf p}_0,\]
  where
  \[\widehat{M}_\alpha=\alpha \widetilde{M}+(1-\alpha)\frac{1}{n}\] and
  \({\bf p}_0=(p_{0}^{(1)},...,p_{0}^{(n)})^T\) is some (actually, any)
  initial vector such that \(p_0^{(i)}\geq 0\) for all \(i\) and that
  \(\sum_{i=1}^n p_{0}^{(i)}=1\).
\item
  For \(\alpha=1\), the \(1\)-PageRank can be defined by the same
  formula as above, \emph{as far as the matrix
  \(\widehat{M}_{1}=\widetilde{M}\) has the eigenvalue \(1\) of the
  multiplicity one}.
\end{enumerate}
\par \vskip 5pt \noindent

The R-code to calculate the \(\alpha\)-PageRank is the following:

\begin{Shaded}
\begin{Highlighting}[]
\NormalTok{PageRank }\OtherTok{\textless{}{-}} \ControlFlowTok{function}\NormalTok{(A,alpha)\{}
\NormalTok{  n }\OtherTok{\textless{}{-}} \FunctionTok{nrow}\NormalTok{(A)}
  \ControlFlowTok{for}\NormalTok{(i }\ControlFlowTok{in} \DecValTok{1}\SpecialCharTok{:}\NormalTok{n)\{A[i,] }\OtherTok{\textless{}{-}}\NormalTok{ A[i,] }\SpecialCharTok{+}\NormalTok{(}\FunctionTok{all}\NormalTok{(A[i,]}\SpecialCharTok{==}\DecValTok{0}\NormalTok{))\}}
\NormalTok{  M }\OtherTok{\textless{}{-}}\NormalTok{ alpha}\SpecialCharTok{*}\FunctionTok{t}\NormalTok{(A)}\SpecialCharTok{\%*\%}\FunctionTok{diag}\NormalTok{(}\DecValTok{1}\SpecialCharTok{/}\FunctionTok{rowSums}\NormalTok{(A),n) }\SpecialCharTok{+}\NormalTok{ (}\DecValTok{1}\SpecialCharTok{{-}}\NormalTok{alpha)}\SpecialCharTok{/}\NormalTok{n}
\NormalTok{  eigen.one }\OtherTok{\textless{}{-}} \FunctionTok{Re}\NormalTok{(}\FunctionTok{eigen}\NormalTok{(M)}\SpecialCharTok{$}\NormalTok{values) }\SpecialCharTok{\textgreater{}} \FloatTok{0.99999}
\NormalTok{  V }\OtherTok{\textless{}{-}} \FunctionTok{Re}\NormalTok{(}\FunctionTok{eigen}\NormalTok{(M)}\SpecialCharTok{$}\NormalTok{vectors[,eigen.one])}
  \ControlFlowTok{if}\NormalTok{(}\FunctionTok{sum}\NormalTok{(eigen.one) }\SpecialCharTok{\textgreater{}} \DecValTok{1}\NormalTok{)\{}
    \FunctionTok{print}\NormalTok{(}\StringTok{"Multiplicity of the eigenvalue 1 is not one"}\NormalTok{)}
\NormalTok{  \}}\ControlFlowTok{else}\NormalTok{\{}
  \FunctionTok{MC\_limit}\NormalTok{(M)[}\DecValTok{1}\SpecialCharTok{:}\NormalTok{n]}
\NormalTok{  \}}
\NormalTok{\}}
\end{Highlighting}
\end{Shaded}

\hypertarget{a-new-statistic-markovrank-for-the-general-non-regular-case}{%
\subsection{A new statistic ``MarkovRank'' (for the general, non-regular
case)}\label{a-new-statistic-markovrank-for-the-general-non-regular-case}}

For any \(\varepsilon \in (0,1]\), let us introduce the
\(((n+1)\times (n+1))\)-matrix \(\check{M}_\varepsilon\) in the
following way; we will actually use this by setting \(\varepsilon=1/k\)
for integers \(k\) in our procedure, namely, \(\check{M}_{1/k}\)’s. 

\textbf{{[}Step 1{]}} Given an \((n \times n)\) adjacency matrix \(A\),
replace each zero row of \(A\) by one row, say, \((1,...,1)\), to create
the new matrix
\(\widetilde{A}=(\widetilde{a}_{ij})_{(i,j) \in \{1,...,n\}^2}\).

\textbf{{[}Step 2{]}} Create the new \(((n+1) \times (n+1))\) matrix
\(\check{A}_\varepsilon=(\check{a}_{ij}^\varepsilon)_{(i,j) \in \{1,...,n+1\}^2}\)
by:
\[\check{a}_{ij}^\varepsilon=\widetilde{a}_{ij}, \quad \forall (i,j) \in \{1,...,n\}^2;\]
\[\check{a}_{i,n+1}^\varepsilon=\varepsilon\sum_{j=1}^n \widetilde{a}_{ij}, \quad \forall i \in \{1,...,n\};\]
\[\check{a}_{n+1,j}^\varepsilon=1,\quad \forall j \in \{1,...,n\}; \quad \check{a}_{n+1,n+1}^\varepsilon=0.\]

\textbf{{[}Step 3{]}} Introduce the \(((n+1) \times (n+1))\) transition
matrix
\[\check{M}_\varepsilon=\check{A}_\varepsilon^T\check{B}_{\varepsilon}^{-1},\]
where \(\check{B}_\varepsilon\) is the \(((n+1) \times (n+1))\) diagonal
matrix whose \((i,i)\) entry is the sum of \(i\)-th row of
\(\check{A}_{\varepsilon}\).

\par \vskip 5pt \noindent
\textbf{Definition 4. {[}The MarkovRank{]}} For any given
\((n \times n)\) adjacency matrix \(A\), the \emph{MarkovRank} is
defined by
\[\mbox{MarkovRank}=\frac{1}{\sum_{i=1}^{n}v_\infty^{(i)}}(v_\infty^{(1)},...,v_\infty^{(n)})^T,\]
where
\({\bf v}_\infty=(v_{\infty}^{(1)},...,v_{\infty}^{(n)},v_{\infty}^{(n+1)})^T\)
is given by
\[{\bf v}_\infty=\lim_{k \to \infty}\check{M}_{1/k}^k{\bf v}_0,\] with
some (actually, any) initial value \({\bf v}_0\) of
\((n+1)\)-dimensional vector
\({\bf v}_0=(v_0^{(1)},...,v_0^{(n)},v_0^{(n+1)})^T\) such that
\(v_{0}^{(i)}\geq 0\) for all \(i\) and that
\(\sum_{i=1}^{n+1}v_0^{(i)}=1\).
\vskip 5pt

The above statistic is implemented in R, as follows:

\begin{Shaded}
\begin{Highlighting}[]
\NormalTok{MarkovRank }\OtherTok{\textless{}{-}} \ControlFlowTok{function}\NormalTok{(A)\{}
\NormalTok{  n }\OtherTok{\textless{}{-}} \FunctionTok{nrow}\NormalTok{(A)}
  \ControlFlowTok{for}\NormalTok{(i }\ControlFlowTok{in} \DecValTok{1}\SpecialCharTok{:}\NormalTok{n)\{A[i,] }\OtherTok{\textless{}{-}}\NormalTok{ A[i,] }\SpecialCharTok{+}\NormalTok{(}\FunctionTok{all}\NormalTok{(A[i,]}\SpecialCharTok{==}\DecValTok{0}\NormalTok{))\}}
\NormalTok{  k }\OtherTok{\textless{}{-}} \DecValTok{1}
\NormalTok{  MR }\OtherTok{\textless{}{-}} \FunctionTok{rep}\NormalTok{(}\DecValTok{1}\SpecialCharTok{/}\NormalTok{n,n)}
\NormalTok{  diff }\OtherTok{\textless{}{-}} \DecValTok{1}
  \ControlFlowTok{while}\NormalTok{(diff }\SpecialCharTok{\textgreater{}} \FloatTok{0.0000001}\NormalTok{)\{}
\NormalTok{  MR.pre }\OtherTok{\textless{}{-}}\NormalTok{ MR}
\NormalTok{  Acheck }\OtherTok{\textless{}{-}} \FunctionTok{rbind}\NormalTok{(}\FunctionTok{cbind}\NormalTok{(A,}\FunctionTok{matrix}\NormalTok{(}\FunctionTok{rowSums}\NormalTok{(A)}\SpecialCharTok{/}\NormalTok{k,n,}\DecValTok{1}\NormalTok{)),}
                  \FunctionTok{matrix}\NormalTok{(}\FunctionTok{c}\NormalTok{(}\FunctionTok{rep}\NormalTok{(}\DecValTok{1}\NormalTok{,n),}\DecValTok{0}\NormalTok{),}\DecValTok{1}\NormalTok{,n}\SpecialCharTok{+}\DecValTok{1}\NormalTok{))}
\NormalTok{  M }\OtherTok{\textless{}{-}} \FunctionTok{t}\NormalTok{(Acheck)}\SpecialCharTok{\%*\%}\FunctionTok{diag}\NormalTok{(}\DecValTok{1}\SpecialCharTok{/}\FunctionTok{rowSums}\NormalTok{(Acheck),n}\SpecialCharTok{+}\DecValTok{1}\NormalTok{)}
\NormalTok{  mr }\OtherTok{\textless{}{-}} \FunctionTok{rep}\NormalTok{(}\DecValTok{1}\SpecialCharTok{/}\NormalTok{(n}\SpecialCharTok{+}\DecValTok{1}\NormalTok{),n}\SpecialCharTok{+}\DecValTok{1}\NormalTok{)}
  \ControlFlowTok{for}\NormalTok{(i }\ControlFlowTok{in} \DecValTok{1}\SpecialCharTok{:}\NormalTok{k)\{}
\NormalTok{    mr }\OtherTok{\textless{}{-}}\NormalTok{ M }\SpecialCharTok{\%*\%}\NormalTok{ mr}
\NormalTok{  \}}
\NormalTok{  MR }\OtherTok{\textless{}{-}}\NormalTok{ mr[}\DecValTok{1}\SpecialCharTok{:}\NormalTok{n]}\SpecialCharTok{/}\FunctionTok{sum}\NormalTok{(mr[}\DecValTok{1}\SpecialCharTok{:}\NormalTok{n])}
\NormalTok{  diff }\OtherTok{\textless{}{-}} \FunctionTok{max}\NormalTok{(}\FunctionTok{abs}\NormalTok{(MR}\SpecialCharTok{{-}}\NormalTok{MR.pre))}
\NormalTok{  k }\OtherTok{\textless{}{-}}\NormalTok{ k}\SpecialCharTok{+}\DecValTok{1}
\NormalTok{  \}}
\NormalTok{  MR}
\NormalTok{\}}
\end{Highlighting}
\end{Shaded}

\hypertarget{toy-examples}{%
\section{Toy examples}\label{toy-examples}}

\hypertarget{non-regular-widetildem-with-the-eigenvalue-1-of-multiplicity-one}{%
\subsection{\texorpdfstring{Non-regular \(\widetilde{M}\) with the
eigenvalue \(1\) of multiplicity
one}{Non-regular \textbackslash widetilde\{M\} with the eigenvalue 1 of multiplicity one}}\label{non-regular-widetildem-with-the-eigenvalue-1-of-multiplicity-one}}

Suppose that the adjacency matrix \(A\) is given by:

\begin{verbatim}
##      [,1] [,2] [,3] [,4] [,5]
## [1,]    0    1    1    1    1
## [2,]    0    0    1    0    0
## [3,]    0    0    0    0    0
## [4,]    0    0    0    0    1
## [5,]    0    0    0    1    0
\end{verbatim}

\includegraphics{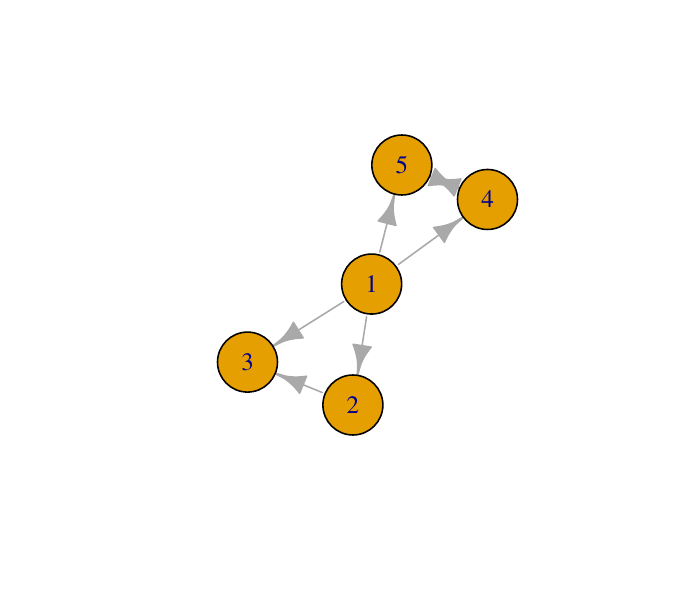}

\begin{itemize}
\tightlist
\item
  The Markov chain generated by the transition matrix \(\widetilde{M}\)
  is not regular; indeed, for any \(k \geq 1\) some entries of the 4th
  and 5th columns of \(\widetilde{M}^k\) are zero.
\item
  However, the multiplicity of the eigenvalue \(1\) of the matrix
  \(\widetilde{M}\) is one. Thus, the intrinsic-PageRank is
  well-defined.
\end{itemize}

Actually, the eigenvalues and eigenvectors of \(\widetilde{M}\) are as
follows:

\begin{Shaded}
\begin{Highlighting}[]
\FunctionTok{eigen}\NormalTok{(}\FunctionTok{M\_tilde}\NormalTok{(A))}
\end{Highlighting}
\end{Shaded}

\begin{verbatim}
## eigen() decomposition
## $values
## [1] -1.0000000+0.0000000i  1.0000000+0.0000000i  0.6777330+0.0000000i
## [4] -0.2388665+0.1292988i -0.2388665-0.1292988i
## 
## $vectors
##                  [,1]             [,2]          [,3]                    [,4]
## [1,] -1.390248e-17+0i  3.431396e-18+0i  0.1798040+0i  0.49993367+0.27061485i
## [2,] -8.890445e-19+0i -5.490233e-17+0i  0.2461296+0i  0.21383811-0.16747727i
## [3,]  5.134079e-17+0i  3.500023e-16+0i  0.6092956+0i -0.77203783+0.00000000i
## [4,] -7.071068e-01+0i  7.071068e-01+0i -0.5176146+0i  0.02913303-0.05156879i
## [5,]  7.071068e-01+0i  7.071068e-01+0i -0.5176146+0i  0.02913303-0.05156879i
##                         [,5]
## [1,]  0.49993367-0.27061485i
## [2,]  0.21383811+0.16747727i
## [3,] -0.77203783+0.00000000i
## [4,]  0.02913303+0.05156879i
## [5,]  0.02913303+0.05156879i
\end{verbatim}

Let us calculate three Ranks for this data.

\begin{Shaded}
\begin{Highlighting}[]
\FunctionTok{PageRank}\NormalTok{(A,}\FloatTok{0.85}\NormalTok{)  }\CommentTok{\# standard{-}PageRank}
\end{Highlighting}
\end{Shaded}

\begin{verbatim}
## [1] 0.04849124 0.05879563 0.10877194 0.39197059 0.39197059
\end{verbatim}

\begin{Shaded}
\begin{Highlighting}[]
\FunctionTok{PageRank}\NormalTok{(A,}\DecValTok{1}\NormalTok{)     }\CommentTok{\# intrinsic{-}PageRank}
\end{Highlighting}
\end{Shaded}

\begin{verbatim}
## [1] 6.017267e-08 8.236898e-08 2.039050e-07 4.999998e-01 4.999998e-01
\end{verbatim}

\begin{Shaded}
\begin{Highlighting}[]
\FunctionTok{MarkovRank}\NormalTok{(A)     }\CommentTok{\# MarkovRank}
\end{Highlighting}
\end{Shaded}

\begin{verbatim}
## [1] 0.0001264742 0.0001580828 0.0003161155 0.4996996637 0.4996996637
\end{verbatim}

Observe that

\begin{Shaded}
\begin{Highlighting}[]
\FunctionTok{nrow}\NormalTok{(A)}
\end{Highlighting}
\end{Shaded}

\begin{verbatim}
## [1] 5
\end{verbatim}

\begin{Shaded}
\begin{Highlighting}[]
\FunctionTok{sum}\NormalTok{( }\FunctionTok{rank}\NormalTok{(}\FunctionTok{PageRank}\NormalTok{(A,}\DecValTok{1}\NormalTok{)) }\SpecialCharTok{==} \FunctionTok{rank}\NormalTok{(}\FunctionTok{PageRank}\NormalTok{(A,}\FloatTok{0.85}\NormalTok{)) )}
\end{Highlighting}
\end{Shaded}

\begin{verbatim}
## [1] 5
\end{verbatim}

\begin{Shaded}
\begin{Highlighting}[]
\FunctionTok{sum}\NormalTok{( }\FunctionTok{rank}\NormalTok{(}\FunctionTok{PageRank}\NormalTok{(A,}\DecValTok{1}\NormalTok{)) }\SpecialCharTok{==} \FunctionTok{rank}\NormalTok{(}\FunctionTok{MarkovRank}\NormalTok{(A)))}
\end{Highlighting}
\end{Shaded}

\begin{verbatim}
## [1] 5
\end{verbatim}

We have found that the rank statistics for three Ranks are identical in
this particular example.

\hypertarget{non-regular-widetildem-with-the-eigenvalue-1-of-multiplicity-two}{%
\subsection{\texorpdfstring{Non-regular \(\widetilde{M}\) with the
eigenvalue \(1\) of multiplicity
two}{Non-regular \textbackslash widetilde\{M\} with the eigenvalue 1 of multiplicity two}}\label{non-regular-widetildem-with-the-eigenvalue-1-of-multiplicity-two}}

Suppose that the adjacency matrix \(A\) is given by:

\begin{verbatim}
##      [,1] [,2] [,3] [,4] [,5] [,6]
## [1,]    0    1    1    1    1    1
## [2,]    0    0    1    1    0    0
## [3,]    0    1    0    1    0    0
## [4,]    0    1    1    0    0    0
## [5,]    0    0    0    0    0    1
## [6,]    0    0    0    0    1    0
\end{verbatim}

\includegraphics{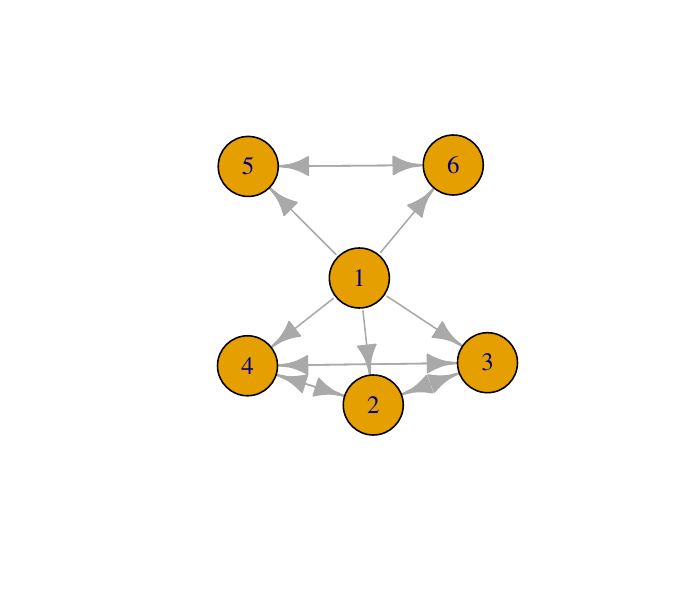}

The multiplicity of the eigenvalue \(1\) of the matrix
\(\widetilde{M}=M\) is two. The Markov chain generated by the transition
matrix \(\widetilde{M}\) is necessarily non-regular.

Actually, the eigenvalues and eigenvectors of \(\widetilde{M}\) are as
follows:

\begin{Shaded}
\begin{Highlighting}[]
\FunctionTok{eigen}\NormalTok{(}\FunctionTok{M\_tilde}\NormalTok{(A))}
\end{Highlighting}
\end{Shaded}

\begin{verbatim}
## eigen() decomposition
## $values
## [1]  1.0 -1.0  1.0 -0.5 -0.5  0.0
## 
## $vectors
##           [,1]       [,2]      [,3]       [,4]       [,5]       [,6]
## [1,] 0.0000000  0.0000000 0.0000000  0.0000000  0.0000000  0.9128709
## [2,] 0.0000000  0.0000000 0.5773503  0.4082483 -0.5661385 -0.1825742
## [3,] 0.0000000  0.0000000 0.5773503 -0.8164966 -0.2264554 -0.1825742
## [4,] 0.0000000  0.0000000 0.5773503  0.4082483  0.7925939 -0.1825742
## [5,] 0.7071068 -0.7071068 0.0000000  0.0000000  0.0000000 -0.1825742
## [6,] 0.7071068  0.7071068 0.0000000  0.0000000  0.0000000 -0.1825742
\end{verbatim}

Let us calculate standard-PageRank, \(0.9999\)-PageRank which is very
close to ``intrinsic-PageRank'', and MarkovRank.

\begin{Shaded}
\begin{Highlighting}[]
\FunctionTok{PageRank}\NormalTok{(A,}\FloatTok{0.85}\NormalTok{)    }\CommentTok{\# standard{-}PageRank}
\end{Highlighting}
\end{Shaded}

\begin{verbatim}
## [1] 0.025 0.195 0.195 0.195 0.195 0.195
\end{verbatim}

\begin{Shaded}
\begin{Highlighting}[]
\FunctionTok{PageRank}\NormalTok{(A,}\FloatTok{0.9999}\NormalTok{)  }\CommentTok{\# very close to intrinsic{-}PageRank}
\end{Highlighting}
\end{Shaded}

\begin{verbatim}
## [1] 1.666667e-05 1.999967e-01 1.999967e-01 1.999967e-01 1.999967e-01
## [6] 1.999967e-01
\end{verbatim}

\begin{Shaded}
\begin{Highlighting}[]
\FunctionTok{PageRank}\NormalTok{(A,}\DecValTok{1}\NormalTok{)       }\CommentTok{\# intrinsic{-}PageRank is not well{-}defined}
\end{Highlighting}
\end{Shaded}

\begin{verbatim}
## [1] "Multiplicity of the eigenvalue 1 is not one"
\end{verbatim}

\begin{Shaded}
\begin{Highlighting}[]
\FunctionTok{MarkovRank}\NormalTok{(A)       }\CommentTok{\# MarkovRank}
\end{Highlighting}
\end{Shaded}

\begin{verbatim}
## [1] 0.000128999 0.199974200 0.199974200 0.199974200 0.199974200 0.199974200
\end{verbatim}

\hypertarget{non-trivial-example-for-rank-statistics}{%
\subsection{Non-trivial example for rank
statistics}\label{non-trivial-example-for-rank-statistics}}

Suppose that the adjacency matrix \(A\) is given by:

\begin{verbatim}
##      [,1] [,2] [,3] [,4] [,5] [,6]
## [1,]    0    1    0    1    1    1
## [2,]    1    0    0    0    0    0
## [3,]    0    1    0    0    1    0
## [4,]    0    1    0    0    0    0
## [5,]    0    0    1    1    0    0
## [6,]    0    0    0    0    0    0
\end{verbatim}

\includegraphics{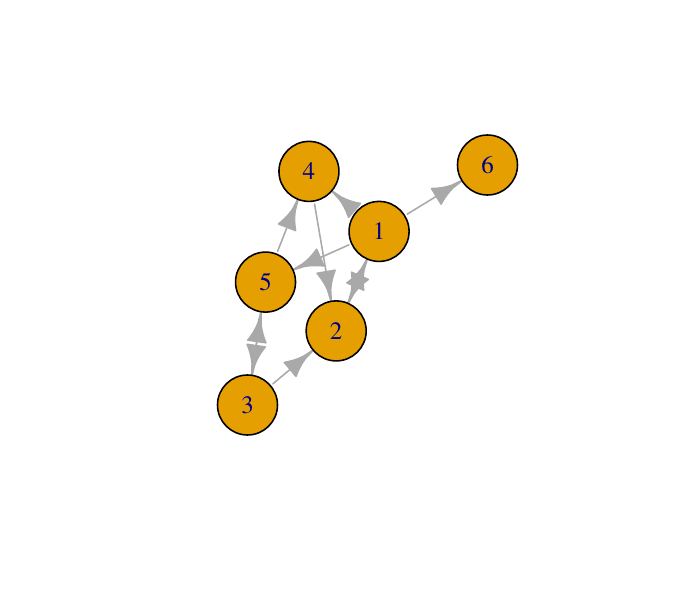}

\begin{Shaded}
\begin{Highlighting}[]
\FunctionTok{PageRank}\NormalTok{(A,}\FloatTok{0.85}\NormalTok{)  }\CommentTok{\# standard{-}PageRank}
\end{Highlighting}
\end{Shaded}

\begin{verbatim}
## [1] 0.26186686 0.26300739 0.09549044 0.15113717 0.13454079 0.09395734
\end{verbatim}

\begin{Shaded}
\begin{Highlighting}[]
\FunctionTok{PageRank}\NormalTok{(A,}\DecValTok{1}\NormalTok{)     }\CommentTok{\# intrinsic{-}PageRank, which is well{-}defined in this example}
\end{Highlighting}
\end{Shaded}

\begin{verbatim}
## [1] 0.28846150 0.27403849 0.07692306 0.14903846 0.12500002 0.08653847
\end{verbatim}

\begin{Shaded}
\begin{Highlighting}[]
\FunctionTok{MarkovRank}\NormalTok{(A)     }\CommentTok{\# MarkovRank}
\end{Highlighting}
\end{Shaded}

\begin{verbatim}
## [1] 0.28832612 0.27398783 0.07701940 0.14904773 0.12505010 0.08656882
\end{verbatim}

Observe an interesting result as follows:

\begin{Shaded}
\begin{Highlighting}[]
\FunctionTok{rank}\NormalTok{(}\FunctionTok{PageRank}\NormalTok{(A,}\FloatTok{0.85}\NormalTok{))}
\end{Highlighting}
\end{Shaded}

\begin{verbatim}
## [1] 5 6 2 4 3 1
\end{verbatim}

\begin{Shaded}
\begin{Highlighting}[]
\FunctionTok{rank}\NormalTok{(}\FunctionTok{PageRank}\NormalTok{(A,}\DecValTok{1}\NormalTok{))}
\end{Highlighting}
\end{Shaded}

\begin{verbatim}
## [1] 6 5 1 4 3 2
\end{verbatim}

\begin{Shaded}
\begin{Highlighting}[]
\FunctionTok{rank}\NormalTok{(}\FunctionTok{MarkovRank}\NormalTok{(A))}
\end{Highlighting}
\end{Shaded}

\begin{verbatim}
## [1] 6 5 1 4 3 2
\end{verbatim}

Let us count the number of the same ranks:

\begin{Shaded}
\begin{Highlighting}[]
\FunctionTok{sum}\NormalTok{( }\FunctionTok{rank}\NormalTok{(}\FunctionTok{PageRank}\NormalTok{(A,}\DecValTok{1}\NormalTok{)) }\SpecialCharTok{==} \FunctionTok{rank}\NormalTok{(}\FunctionTok{PageRank}\NormalTok{(A,}\FloatTok{0.85}\NormalTok{)) )}
\end{Highlighting}
\end{Shaded}

\begin{verbatim}
## [1] 2
\end{verbatim}

\begin{Shaded}
\begin{Highlighting}[]
\FunctionTok{sum}\NormalTok{( }\FunctionTok{rank}\NormalTok{(}\FunctionTok{PageRank}\NormalTok{(A,}\DecValTok{1}\NormalTok{)) }\SpecialCharTok{==} \FunctionTok{rank}\NormalTok{(}\FunctionTok{MarkovRank}\NormalTok{(A)) )}
\end{Highlighting}
\end{Shaded}

\begin{verbatim}
## [1] 6
\end{verbatim}

We have found that, in this example, although the rank statistics of
standard-PageRank and intrinsic-PageRank are different, those of
intrinsic-PageRank and MarkovRank are identical. In the Appendix, we
shall mathematically prove that the latter half of this observation
\emph{always} holds true.

\hypertarget{a-real-data-analysis}{%
\section{A real data analysis}\label{a-real-data-analysis}}

In this section, we analyze Twitter-following data among US senators as
directed network data. In this data set, an edge represents an instance
of a senator following another senator's Twitter account. The data
consist of two files, one listing pairs of the Twitter screen names of
the following and followed politicians, \texttt{twitter-following.csv},
and the other containing information about each politician,
\texttt{twitter-senator.csv}.

\begin{Shaded}
\begin{Highlighting}[]
\NormalTok{twitter }\OtherTok{\textless{}{-}} \FunctionTok{read.csv}\NormalTok{(}\StringTok{"twitter{-}following.csv"}\NormalTok{, }\AttributeTok{stringsAsFactors=}\ConstantTok{FALSE}\NormalTok{)}
\NormalTok{senator }\OtherTok{\textless{}{-}} \FunctionTok{read.csv}\NormalTok{(}\StringTok{"twitter{-}senator.csv"}\NormalTok{, }\AttributeTok{stringsAsFactors=}\ConstantTok{FALSE}\NormalTok{)}
\end{Highlighting}
\end{Shaded}

Let us create the adjacency matrix \(A\) for this network.

\begin{Shaded}
\begin{Highlighting}[]
\NormalTok{n }\OtherTok{\textless{}{-}} \FunctionTok{nrow}\NormalTok{(senator)}
\NormalTok{twitter.adj }\OtherTok{\textless{}{-}} \FunctionTok{matrix}\NormalTok{(}\DecValTok{0}\NormalTok{, }\AttributeTok{nrow=}\NormalTok{n, }\AttributeTok{ncol=}\NormalTok{n)}
\FunctionTok{rownames}\NormalTok{(twitter.adj) }\OtherTok{\textless{}{-}}\NormalTok{ senator}\SpecialCharTok{$}\NormalTok{screen\_name}
\FunctionTok{colnames}\NormalTok{(twitter.adj) }\OtherTok{\textless{}{-}}\NormalTok{ senator}\SpecialCharTok{$}\NormalTok{screen\_name}
\ControlFlowTok{for}\NormalTok{(i }\ControlFlowTok{in} \DecValTok{1}\SpecialCharTok{:}\FunctionTok{nrow}\NormalTok{(twitter))\{}
\NormalTok{  twitter.adj[twitter}\SpecialCharTok{$}\NormalTok{following[i],}
\NormalTok{              twitter}\SpecialCharTok{$}\NormalTok{followed[i]] }\OtherTok{\textless{}{-}} \DecValTok{1}
\NormalTok{\}}
\end{Highlighting}
\end{Shaded}

Let us observe the in-degrees of the nodes:

\scriptsize

\begin{Shaded}
\begin{Highlighting}[]
\FunctionTok{colSums}\NormalTok{(twitter.adj)   }\CommentTok{\# in{-}degree}
\end{Highlighting}
\end{Shaded}

\begin{verbatim}
##    SenAlexander        RoyBlunt    SenatorBoxer SenSherrodBrown     SenatorBurr 
##              52              57              44              45              54 
##  SenatorBaldwin     JohnBoozman SenJohnBarrasso     SenBennetCO   SenBlumenthal 
##              37              47              55              43              44 
##    SenBookerOfc SenatorCantwell   SenatorCardin   SenatorCarper     SenDanCoats 
##               4              37              46              44              56 
##  SenThadCochran       MikeCrapo  SenatorCollins       SenCapito      JohnCornyn 
##              44              49              58              28              55 
##     SenBobCasey    SenBobCorker  SenCoonsOffice    SenTomCotton      SenTedCruz 
##              43              55              30              31              34 
##   SenatorDurbin     SenDonnelly     SteveDaines     SenatorEnzi    SenJoniErnst 
##              47              37              27              42              30 
##    SenFeinstein       JeffFlake  SenatorFischer   ChuckGrassley  SenCoryGardner 
##              40              41              40              55              30 
##   SenOrrinHatch   SenDeanHeller     maziehirono  MartinHeinrich   SenJohnHoeven 
##              44              55              39              34              48 
## SenatorHeitkamp     InhofePress  SenatorIsakson   SenRonJohnson     SenatorKirk 
##              41              42              49              56              53 
##    SenAngusKing  SenKaineOffice    SenatorLeahy SenatorLankford      SenMikeLee 
##              37              13              49              31              43 
##   SenJohnMcCain  McConnellPress SenatorMenendez     SenatorBarb      JerryMoran 
##              64              41              49              47              52 
##     PattyMurray   lisamurkowski   ChrisMurphyCT McCaskillOffice  SenJeffMerkley 
##              45              60              35              33              46 
##  Sen_JoeManchin   SenBillNelson   senrobportman   SenGaryPeters  sendavidperdue 
##              51              46              58              25              30 
##     SenJackReed     SenatorReid   SenPatRoberts    SenatorRisch   SenRubioPress 
##              41              52              49              38              43 
##   SenatorRounds      SenSanders      SenSchumer       SenShelby     SenStabenow 
##              31              44              46              41              40 
## SenatorSessions  SenatorShaheen SenatorTimScott  SenBrianSchatz        SenSasse 
##              49              44              41              29              27 
##  SenDanSullivan    SenJohnThune       SenToomey   SenatorTester   SenThomTillis 
##              17              52              58              23              28 
## SenatorTomUdall   SenatorWicker        RonWyden   SenWhitehouse      MarkWarner 
##              46              47              48              45              44 
##       SenWarren 
##              39
\end{verbatim}

\normalsize

Let us finish up this paper with observing the results of three Ranks
for this network data.

\begin{Shaded}
\begin{Highlighting}[]
\NormalTok{standard\_PR.result }\OtherTok{\textless{}{-}} \FunctionTok{PageRank}\NormalTok{(twitter.adj,}\FloatTok{0.85}\NormalTok{)}
\NormalTok{intrinsic\_PR.result }\OtherTok{\textless{}{-}} \FunctionTok{PageRank}\NormalTok{(twitter.adj,}\DecValTok{1}\NormalTok{)}
\NormalTok{MR.result }\OtherTok{\textless{}{-}} \FunctionTok{MarkovRank}\NormalTok{(twitter.adj)}
\FunctionTok{names}\NormalTok{(standard\_PR.result) }\OtherTok{\textless{}{-}}\NormalTok{ senator}\SpecialCharTok{$}\NormalTok{screen\_name}
\FunctionTok{names}\NormalTok{(intrinsic\_PR.result) }\OtherTok{\textless{}{-}}\NormalTok{ senator}\SpecialCharTok{$}\NormalTok{screen\_name}
\FunctionTok{names}\NormalTok{(MR.result) }\OtherTok{\textless{}{-}}\NormalTok{ senator}\SpecialCharTok{$}\NormalTok{screen\_name}
\end{Highlighting}
\end{Shaded}

The top six Twitter accounts, by the three Ranks, are given as follows.
Although the Rank values for three definitions are somewhat different,
the rankings of top and last Twitter accounts are identical.

\begin{Shaded}
\begin{Highlighting}[]
\FunctionTok{head}\NormalTok{(}\FunctionTok{sort}\NormalTok{(standard\_PR.result,  }\AttributeTok{decreasing=}\ConstantTok{TRUE}\NormalTok{))}
\end{Highlighting}
\end{Shaded}

\begin{verbatim}
##  SenJohnMcCain     JohnCornyn MartinHeinrich  lisamurkowski      SenToomey 
##     0.02225510     0.01994213     0.01945448     0.01873310     0.01721254 
##    SenDanCoats 
##     0.01654421
\end{verbatim}

\begin{Shaded}
\begin{Highlighting}[]
\FunctionTok{head}\NormalTok{(}\FunctionTok{sort}\NormalTok{(intrinsic\_PR.result, }\AttributeTok{decreasing=}\ConstantTok{TRUE}\NormalTok{))}
\end{Highlighting}
\end{Shaded}

\begin{verbatim}
##  SenJohnMcCain     JohnCornyn MartinHeinrich  lisamurkowski      SenToomey 
##     0.02441628     0.02196977     0.02149121     0.02031664     0.01846398 
##    SenDanCoats 
##     0.01762956
\end{verbatim}

\begin{Shaded}
\begin{Highlighting}[]
\FunctionTok{head}\NormalTok{(}\FunctionTok{sort}\NormalTok{(MR.result, }\AttributeTok{decreasing=}\ConstantTok{TRUE}\NormalTok{))}
\end{Highlighting}
\end{Shaded}

\begin{verbatim}
##  SenJohnMcCain     JohnCornyn MartinHeinrich  lisamurkowski      SenToomey 
##     0.02437806     0.02193313     0.02145419     0.02028841     0.01844162 
##    SenDanCoats 
##     0.01761033
\end{verbatim}

However, once we observe the rankings for the whole accounts, we again
find that the rank statistic of standard-PageRank does not coincide with
that of intrinsic-PageRank, but that of MarkovRank does.

\begin{Shaded}
\begin{Highlighting}[]
\FunctionTok{length}\NormalTok{(intrinsic\_PR.result)}
\end{Highlighting}
\end{Shaded}

\begin{verbatim}
## [1] 91
\end{verbatim}

\begin{Shaded}
\begin{Highlighting}[]
\FunctionTok{sum}\NormalTok{( }\FunctionTok{rank}\NormalTok{(intrinsic\_PR.result) }\SpecialCharTok{==} \FunctionTok{rank}\NormalTok{(standard\_PR.result) )}
\end{Highlighting}
\end{Shaded}

\begin{verbatim}
## [1] 46
\end{verbatim}

\begin{Shaded}
\begin{Highlighting}[]
\FunctionTok{sum}\NormalTok{( }\FunctionTok{rank}\NormalTok{(intrinsic\_PR.result) }\SpecialCharTok{==} \FunctionTok{rank}\NormalTok{(MR.result) )}
\end{Highlighting}
\end{Shaded}

\begin{verbatim}
## [1] 91
\end{verbatim}

\hypertarget{appendix-main-theorem-and-its-proof}{%
\section*{Appendix: Main theorem and its
proof}\label{appendix-main-theorem-and-its-proof}}

In the main context of this paper, we have confirmed that:

\begin{enumerate}
\def\labelenumi{\arabic{enumi}.}
\item
  The MarkovRank seems to behave as stably as the standard-PageRank,
  while the behavior of the intrinsic-PageRank is not so stable;
\item
  The rank statistic of the standard-PageRank does not always coincide
  with that of the intrinsic-PageRank, even when the latter is
  well-defined; 
\item
  The rank statistic of the MarkovRank seems to coincide ``always'' with
  that of the intrinsic-PageRank, as far as the latter is well-defined.
\end{enumerate}

Here, we shall mathematically prove the above
(loose) claim 3 as our main theorem of the current paper.

\par \vskip 5pt \noindent
\textbf{Main Theorem.} If the intrinsic-PageRank corresponding to given
adjacency matrix \(A\) is well-defined, then the rank statistic of
MarkovRank, which is always well-defined, is identical to that of
intrinsic-PageRank.
\par \vskip 5pt \noindent

\par\noindent
\emph{Proof.} As preliminaries, let us introduce the
\(((n+1) \times (n+1))\) transition matrices \(\mathbb{A}\),
\(\mathbb{B}\) and \(\mathbb{C}\) given by:
\begin{eqnarray*}\mathbb{A}&=&
\left(\begin{array}{cccc}
 & & &1/n \\
 &\widetilde{M}& &\vdots \\
 & & &1/n \\
0&\cdots&0&0
\end{array}
\right); \\
\mathbb{B}&=&
\left(\begin{array}{cccc}
 & & &1/n \\
 &O& &\vdots \\
 & & &1/n\\
1&\cdots&1&0
\end{array}
\right);\\
\mathbb{B}^2&=&
\left(\begin{array}{cccc}
1/n&\cdots&1/n&0 \\
\vdots& &\vdots&\vdots \\
1/n&\cdots&1/n&0\\
0&\cdots&0&1
\end{array}
\right). 
\end{eqnarray*} Furthermore, for any \(k=1,2,...\), put
\[\alpha_k=\frac{1}{1+(1/k)}\quad \mbox{and} \quad
\beta_k=\frac{1/k}{1+(1/k)}.\] Then it holds that
\[\check{M}_{1/k}=\alpha_k \mathbb{A} + \beta_k \mathbb{B}.\]

Now, fix any initial vector
\({\bf v}_0=(v_0^{(1)},v_0^{(2)},...,v_0^{(n)},v_0^{(n+1)})^T\) such
that \(v_0^{(i)}\geq 0\) for all \(i\) and that
\(\sum_{i=1}^{n+1}v_0^{(i)}=1\), and for every \(k=1,2,...\), define the
\((n+1)\)-dimensional vectors \({\bf v}_{1}^{(1/k)}\),
\({\bf v}_{2}^{(1/k)}\),\ldots,\({\bf v}_k^{(1/k)}\) by the recursive
formula
\[{\bf v}_{k}^{(1/k)}=\check{M}_{1/k} {\bf v}_{k-1}^{(1/k)}=\cdots=\check{M}_{1/k}^{k-1}{\bf v}_1^{(1/k)}=\check{M}_{1/k}^k {\bf v}_0, \quad k=1,2,....\]

We shall compute \(\check{M}_{1/k}^k\) for general integers \(k\).
First, notice the following facts: \[\mathbb{A}^k=
\left(\begin{array}{cccc}
 & & &m^{k-1,(1)} \\
 &\widetilde{M}^k & &\vdots \\
 & & &m^{k-1,(n)} \\
0&\cdots&0&0
\end{array}
\right),\] where the vector \({\bf m}^{k}=(m^{k,(1)},...,m^{k,(n)})^T\)
is given by
\[m^{k,(i)}=\frac{1}{n}\sum_{l=1}^n \widetilde{M}_{i,l}^{k} \quad \mbox{where} \quad \widetilde{M}_{i,j}^{k} \mbox{ is the }(i,j)\mbox{ entry of the matrix }\widetilde{M}^k,\]
with the convention that
\[\widetilde{M}^{0}=\mathbb{M}_n=\left(\begin{array}{ccc}1/n & \cdots & 1/n \\ \vdots &\cdots & \vdots \\ 1/n & \cdots & 1/n \end{array}\right);\]
\[\mathbb{B}^k=\left\{\begin{array}{cc}\mathbb{B} & \mbox{for odd positive integer }k,\\
                              \mathbb{B}^2 & \mbox{for even positive integer }k;\end{array}\right.\]
\[\mathbb{BA}=\mathbb{C}=\left(\begin{array}{cccc}0&\cdots&0&0\\
                                                          \vdots&\cdots&\vdots&\vdots\\
                                                          0&\cdots&0&0\\
                                                          1&\cdots&1&1\end{array}\right) \quad \mbox{and} \quad \mathbb{B}^2\mathbb{A}=\mathbb{D}=\left(\begin{array}{cccc}1/n&\cdots&1/n&1/n\\
                                                          \vdots&\cdots&\vdots&\vdots\\
                                                         1/n&\cdots&1/n&1/n\\
                                                         0&\cdots&0&0\end{array}\right);\]
\begin{eqnarray*}\mathbb{A}^{k-l}\mathbb{B}^l&=&\left(\begin{array}{cccc}m^{k-l-1,(1)}&\cdots&m^{k-l-1,(1)}&m^{k-l,(1)}\\
                                \vdots & \cdots & \vdots & \vdots \\
                                m^{k-l-1,(n)}&\cdots&m^{k-l-1,(n)}&m^{k-l,(n)}\\
                                0 & \cdots & 0 & 0
                          \end{array}\right), \quad \mbox{for odd }l,\\ \mathbb{A}^{k-l}\mathbb{B}^l&=&\left(\begin{array}{cccc}m^{k-l,(1)}&\cdots&m^{k-l,(1)}&m^{k-l-1,(1)}\\
                                     \vdots & \cdots & \vdots & \vdots \\
                                     m^{k-l,(n)}&\cdots&m^{k-l,(n)}&m^{k-l-1,(n)}\\
                                     0 & \cdots & 0 & 0
                          \end{array}\right),\quad \mbox{for even }l;\end{eqnarray*}
\[\mathbb{CA}=\mathbb{CB}=\mathbb{C} \quad \mbox{and}\quad \mathbb{DA}=\mathbb{DB}=\mathbb{D}.\]
Hence it holds for any \(k\geq 3\) that
\[\check{M}_{1/k}^k=\alpha_k^k\mathbb{A}^k+\sum_{l=1}^{k-1}\alpha_k^{k-l}\beta_{k}^{l} \mathbb{A}^{k-l}\mathbb{B}^{l}+\beta_k^k \mathbb{B}^k +\gamma_{k}\mathbb{C}+\delta_{k}\mathbb{D},\]
where \(\gamma_{k},\delta_{k}\geq 0\) and
\(\sum_{l=0}^{k}\alpha_{k}^{k-l}\beta_{k}^{l}+\gamma_{k}+\delta_{k}=1\),
with \(\delta_k=\alpha_k\gamma_k\).

Notice that \(\lim_{k \to \infty}\alpha_k^k=1/e\). As for the second
term on the right-hand side, we have that any entry is non-negative and
bounded by
\[\sum_{l=1}^{k-1}\alpha_k^{k-l}\beta_{k}^{l}\leq \alpha_k^k \sum_{l=1}^\infty(\beta_k/\alpha_k)^{l}\leq\frac{1/k}{1-(1/k)},\]
which converges to zero as \(k \to \infty\). It is clear that the third
term also vanishes as \(k \to \infty\). As for the last two terms, since
\(\lim_{k \to \infty}(\gamma_k+\delta_k)=1-(1/e)\) and
\(\delta_k=\alpha_k\gamma_k\) for \(k\geq 3\), we have that
\(\lim_{k \to \infty}\gamma_k=\lim_{k \to \infty}\delta_k=(1-(1/e))/2\).
Hence, it holds
that\[\check{\mathbb{M}}=\lim_{k \to 0}\check{M}_{1/k}^k=(1/e)\mathbb{A}^\infty+\frac{1-(1/e)}{2}(\mathbb{C}+\mathbb{D}).\]
Note that the multiplicity of the eigenvalue \(1\) for this matrix is
one. Thus, when the multiplicity of the eigenvalue \(1\) for the matrix
\(\widetilde{M}\) is one, the rescaled eigenvector of \((n \times n)\)
matrix \(\widetilde{M}^\infty\) and the first \(n\) entries of the rescaled
eigenvector of \(((n+1) \times (n+1))\) matrix \(\check{\mathbb{M}}\),
both for the eigenvalue \(1\), have the same rank statistics. The
assertion of the theorem has been proved.

\vskip 20pt
\par\noindent
\textbf{Acknowledgments:} I thank Professors Satoshi Inaba and Kosuke
Imai for helpful comments and discussions. Professor Imai kindly gave me
a permission to use the data sets from the accompanying web-site of his
book {[}4{]}.

\hypertarget{references-and-remark}{%
\section*{References and remark}\label{references-and-remark}}

\vskip 5pt

\par

\noindent
[1] Bar-Yossef, Z. and Mashiach, L. (2008). Local Approximation of
PageRank and Reverse PageRank. In: Proceedings of the 31st ACM SIGIR
Conference on Research and Development Information Retrieval (CIGIR
2008), July 20-24, 2008, Singapore.

\vskip 5pt

\par

\noindent
[2] Bianchini, M., Gori, M. and Scarselli, F. (2005). Inside PageRank.
\emph{ACM Trans. Internet Technol.} \textbf{5}, No.1, 92-128.

\vskip 5pt

\par

\noindent
[3] Brin, S. and Page, L. (1998). The Anatomy of a Large-Scale
Hypertextual Web Search Engine. In: Seventh International World-Wide Web
Conference (WWW 1998), April 14-18, 1998, Brisbane, Australia.

\vskip 5pt

\par

\noindent
[4] Imai, K. (2017). \emph{Quantitative Social Science: An
Introduction.} Princeton University Press.

\vskip 5pt

\par

\noindent
[5] Langville, A.N. and Meyer, C.D.D. (2011). \emph{Google's PageRank
and Beyond: The Science of Search Engine Rankings.} Princeton University
Press. Princeton.

\vskip 20pt
\par\noindent
\textbf{Remark:} The data files \texttt{twitter-following.csv} and
\texttt{twitter-senator.csv} treated in Section 5 are available on the
accompanying web-site for the book {[}4{]}:
\texttt{http://qss.princeton.press/}

\end{document}